\documentstyle[12pt,epsfig]{article}

\newcommand{\mathbold}[1]{\mbox{\rm\bf #1}}

\newcommand{\beq}{\begin{equation}}
\newcommand{\eeq}{\end{equation}}
\newcommand{\nn}{\nonumber}
\newcommand{\bea}{\begin{eqnarray}}
\newcommand{\eea}{\end{eqnarray}}

\newcommand{\gtrsim}{\ \rlap{\raise 2pt\hbox{$>$}}{\lower 2pt \hbox{$\sim$}}\ }
\newcommand{\lessim}{\ \rlap{\raise 2pt\hbox{$<$}}{\lower 2pt \hbox{$\sim$}}\ }

\newcommand{\np}[1]{Nucl. Phys. {\bf #1}}
\newcommand{\pl}[1]{Phys. Lett. {\bf #1}}
\newcommand{\pr}[1]{Phys. Rev. {\bf #1}}
\newcommand{\prl}[1]{Phys. Rev. Lett. {\bf #1}}

\makeatletter
\if@twoside
\else \oddsidemargin 00pt \evensidemargin 00pt
\fi
\let\@eqnsel = \hfil

\expandafter\ifx\csname mathrm\endcsname\relax\def\mathrm#1{{\rm #1}}\fi
\@ifundefined{mathrm}{\def\mathrm#1{{\rm #1}}}{\relax}

\makeatother

\begin{document}
\thispagestyle{empty}
\null
\hfill FTUV/96-24,IFIC/96-26

\hfill hep-ph/9608450

\vskip 1.5cm

\begin{center}
{\Large \bf      %INSERT TITLE
 Spontaneous $CP$-violation in the left-right model
and the kaon system
\par} \vskip 2.em
{\large		%INSERT NAMES
{\sc G. Barenboim$^{1,2}$, J. Bernab\'eu$^1$ and M. Raidal$^{1,2}$
}  \\[1ex] %INSERT ADDRESS
{\it $^1$ Departament de F\'\i sica Te\`orica, Universitat 
de Val\`encia}\\ 
{\it $^2$ IFIC, Centre Mixte Universitat 
de Val\`encia - CSIC} \\
{\it E-46100 Burjassot, Valencia, Spain} \\[1ex]
\vskip 0.5em
\par} 
\end{center} \par
\vfil
{\bf Abstract} \par

A left-right model with spontaneous $CP$ breakdown,  consistent 
with the particle physics phenomenology, is presented. 
Constraints on  free parameters of the model: mass of the new
right handed gauge boson $M_2$ and  ratio $r$ of the two vacuum expectation
values of the bidoublet, are found
from the measurement of  $\epsilon$ in the kaon system.
For most of the parameter space, $M_2$ is restricted  to be below 10 TeV.
Higher masses can be achieved only by fine tuning of 
Kobayashi-Maskawa matrix elements, quark masses, $r$ and the phase
$\alpha$ which is the unique source of $CP$-violation in the model. 
Large number of  combinations of signs of quark masses,
 which are observables of the model, are found to be not allowed 
since they contradict with data.   
The range of $\epsilon'/\epsilon$ the model predicts is around $10^{-4}$
in magnitude.

\par
\vskip 0.5cm
\noindent July 1996 \par
\null
\setcounter{page}{0}
\clearpage

\section{Introduction}

The origin of $CP$-violation  \cite{jarls1} remains a mystery despite of the
spectacular progress made during the last twenty years in 
understanding the weak and electromagnetic interactions in the framework
of 
spontaneously broken gauge theories.
The Standard Model (SM) allows, in its six-quark version, for the 
appearance of a phase in the effective quark-quark-vector boson
vertex \cite{KM}.
This phase, the Kobayashi-Maskawa (KM) phase, can be used to parametrize
the amount of $CP$-violation in the SM.
More precisely, this phase is responsible in the kaon system 
for a non-vanishing value of
the $\epsilon$ paramete,r which measures  the amount of
$\Delta S = 2$ $CP$-violation. In this model, 
the parameter $\epsilon^\prime$, which measures the $\Delta S = 1 $
amount of $CP$-violation, turns out to be naturally small \cite{epsp},
in agreement with the present  experimental result \cite{pdb}.

However, the KM mechanism for incorporating $CP$-violation to the SM
cannot be fully satisfactory since it does not explain where  
$CP$-violation comes from.
Moreover, there are indications that the
amount of  $CP$-violation  in the SM 
is not sufficient for generating  baryon asymmetry of the Universe \cite{uni}.
Therefore one has to look for possible
 sources of  $CP$-violation beyond the SM.  

One of the most attractive extensions of the SM is the model based on
the gauge   group $SU(2)_L \times SU(2)_R \times U(1)_{B-L} $ \cite{lr}.
In addition to the original idea of providing an explanation
to  the observed parity violation of the
weak interaction at low energies it also  turned  out to be capable  
to explain  the lightness of the ordinary neutrinos  
via the so-called see-saw mechanism \cite{ss}.  
In this model the Lagrangian is left-right symmetric but the vacuum
is not invariant under the parity transformation. The left-right 
symmetry is spontaneously broken.  At high energies 
the new particle degrees of freedom like the new right handed gauge 
bosons $W_2,$ will appear.

The same argumentation can be applyed in the case of $CP$-violation. 
One can assume that the original Lagrangian is symmetric under $CP$ 
transformation but the vacuum breaks $CP$ spontaneously.
Despite of the fact that the processes with spontaneous $CP$-violation 
in the left-right model have been studied previously in several works 
\cite{chang,eg,frere},
a careful analysis of left-right models indicated that it could be 
impossible to
construct a phenomenologically acceptable  model with  spontaneously
broken $CP$ \cite{nolr}
 because of the flavour changing neutral currents (FCNC)
occurring in those models.
However, it turned out that not all the solutions had been taken 
into account in these  analysis.              
It was shown in Ref.\cite{wu} that in two-doublet models  $CP$-violation 
can occur spontaneously without violating FCNC restrictions. Recently,
a similar result  was shown  for the left-right models \cite{gabriela}.
Namely, even with the minimal Higgs sector containing a bidoublet 
$\phi$ and 
two triplets $\Delta_{L,R}$
it is possible to obtain spontaneous breakdown of $CP$ and
satisfy FCNC constraints. 
For this issue the $\beta$ terms of the Higgs potential, the  
non diagonal quartic couplings between the two scalar triplets and
the bidoublet,
which were taken to be zero in previous works, play a crucial role.

Motivated by these  results we  re-analyse the 
$K^0$-$\bar K^0$ system assuming a phenomenologically
consistent  left-right model   with a discrete $CP$ symmetry at 
the Lagrangian level, i.e. $CP$ is violated only spontaneously. 
All  $CP$-violating observables in the SM are proportional 
to $\lambda^6$ in the Wolfenstein parametrization.
 However, since $CP$-violation in the left-right model can occur 
even with two quark generations
the dominant left-right contribution in the kaon system 
is proportional to $\lambda^2$ only.  
This makes the kaon system very sensitive to searches for 
the left-right symmetry.
In our analysis we take into account   new  measurements of the quark masses,
  KM  matrix elements  and  strong coupling constant $\alpha_s$
 as well as the recent 
developments in understanding of hadronic matrix elements and QCD 
corrections in the left-handed \cite{bur} and right-handed \cite{frere}
sectors of the $K$ system. 
 We show that with the present
data the measurements of $\epsilon$ and $\epsilon'/\epsilon$
allow us to restrict the parameter space of the model considerably. 
In particular, without fine tuning, the ratio $r$ of 
vacuum expectation values (vevs) of the bidoublet and 
the mass of the new right handed gauge boson $M_2$ should be limited 
to a quite narrow range in order to explain the observed $CP$-violation.

The outline of the paper is the following. In Section 2 we present our model
in detail.  In Section 3 we parametrize the most general KM matrix in
terms of a single $CP$-violating phase arising from the spontaneos symmetry
breaking (SSB). 
In order to do that
we  study the quark mass matrices in the model. In Section 4
we carry out the analysis of $\epsilon$ and $\epsilon'/\epsilon$
 in terms of our model  and find restrictions on the model parameters. 
  A summary is given in Section 5.

\section{Left-right symmetric model with spontaneous $CP$-violation}

We begin with presenting  the minimal $SU(2)_L \times SU(2)_R 
\times U(1)_{B-L} $ model with a left-right discrete symmetry. 
In the left-right symmetric models each generation of quarks and
leptons are assigned to the multiplets
\bea
Q = \pmatrix { u \cr d \cr} \; \; , \; \; L=\pmatrix{ \nu \cr e \cr} \;\; ,
\eea
with the quantum numbers $(T_L,T_R,B-L)$
\bea
Q_L \; :\; \left( \frac{1}{2},0,\frac{1}{3} \right) \;\;\;  , \;\;\; 
L_L \; :\; \left( \frac{1}{2},0,-1 \right) , \nn \\
Q_R \; :\; \left( 0,\frac{1}{2},\frac{1}{3} \right)  \;\;\;  , \;\;\; 
L_R \; :\; \left(0,\frac{1}{2},-1 \right) .
\eea
$CP$-violation in the model will arise from the Higgs sector and so
we must spend a bit of time for a more detailed description of  this sector.

The Higgs sector consists of a bidoublet
\bea
\phi = \pmatrix {\phi_1^0 & \phi_1^+ \cr \phi_2^- & \phi_2^0 \cr}
\eea
and the triplets
\bea
\Delta_{L,R} = \pmatrix {\frac{\Delta_{L,R}^+}{\sqrt{2}}  & 
\Delta_{_{L,R}}^{^{++}} \cr 
\Delta_{_{L,R}}^{^0} & \frac{-\Delta_{L,R}^+}{\sqrt{2}}\cr}
\eea
with the quantum numbers $
 \phi\;: \; (\frac{1}{2},\frac{1}{2}^\ast,0) \;\; , \;\; 
\Delta_L \;: \; (1,0,2) \;\;, \;\; \Delta_R \; : \; (0,1,2) , $
respectively.

Only the fields $\phi_1^0$, $\phi_2^0$, $\Delta_L^0$ and 
$\Delta_R^0$ can acquire vevs without 
violating  electric charge. If $\Delta_L$ or $\Delta_R$ acquire a vev,
 then $B-L$ is necessarily broken.
Further, if $\langle \Delta_R \rangle \neq \langle \Delta_L 
\rangle $, then parity breakdown is also ensured. The new feature
that we have analysed, and we want to discuss in this work, is
the phenomenological consequence of supposing that the
 vevs  of the neutral fields are not real.
In this case $CP$ is spontaneously broken and we arrive at a unified picture 
of parity, time-reversal and $B-L$ violation.

In general, our symmetry breaking would be triggered by the  vevs
\bea
\langle \phi \rangle  = 
\pmatrix {\frac{k_1}{\sqrt{2}} & 0 \cr 0 & \frac{k_2}{\sqrt{2}}\cr}
\; \; \; \; , \; \; \; \;
\langle \Delta_{L,R} \rangle = \pmatrix {0   & 0 \cr 
\frac{v_{L,R}}{\sqrt{2}} & 0 \cr},
\eea
satisfying the following hierarchy: $ |v_R| \gg |k_1|, |k_2|  \gg |v_L| $.
Here all the vevs can be complex. However, we still have a freedom
 to absorb two of these phases
in such a way that two vevs are real and two complex:
 $ v_R=|v_R| e^{i \theta}$ and  $k_2=|k_2| e^{i\alpha}.$

The Higgs sector contains 20 degrees of freedom of which 14 correspond
to physical states, the latter split into four doubly-charged,
four singly-charged and six neutral scalar fields.
The remaining six degrees of freedom are eaten by the massive gauge bosons 
$W_{1,2}^{\pm}$ and $Z^0_{1,2}$ during the SSB.

Let us now discuss the form of the scalar field potential. The
discrete left-right symmetry requires the potential to
be invariant under
\beq
\Psi_L \longleftrightarrow \Psi_R \;\;\; 
\Delta_L \longleftrightarrow \Delta_R  \;\;\;
\phi \longleftrightarrow \phi^\dagger,
\label{trans}
\eeq
where $\Psi$ denotes any fermion.
We assume that the global 
phases allowed to appear in the transformations above  are absorbed by
the proper redefinition of the fields.
Further, the most general scalar field potential cannot have trilinear
terms: because of the  nonzero $B-L$ quantum numbers of the $\Delta_L$ and
$\Delta_R$ triplets, these must always appear in the quadratic combinations 
$\Delta_L^\dagger \Delta_L $, $\Delta_R^\dagger \Delta_R $,
$\Delta_L^\dagger \Delta_R $ or $\Delta_R^\dagger \Delta_L $.
These combinations can never be combined with a single bidoublet $\phi$ 
in such a way as to form  $SU(2)_L$ and $SU(2)_R$ singlets.
Nor can three bidoublets be combined so as to yield a singlet.
However, quartic combinations of the form $ \beta \;
\mbox{Tr} ( \Delta_L^\dagger \phi
\Delta_R \phi^\dagger )$  are  allowed by the left-right symmetry.
Following these conditions the most general form of the 
Higgs potential is 
\beq
V=V_{\phi}+V_{\Delta}+V_{\phi\Delta},
\eeq
where
\bea
{\mathbold V}_\phi & = & -\mu^2_{i,j}  \mbox{Tr}(\phi^\dagger_i \phi_j ) \, 
+ \,
 \lambda_{i,j,k,l} \,  \mbox{Tr}( \phi^\dagger_i \phi_j )  
 \, \mbox{Tr}(\phi^\dagger_k  \phi_l) ,
\nn \eea

\bea
{\mathbold V}_\Delta  & = & -\mu^2_i  \, \mbox{Tr}(\Delta_i \Delta_i^\dagger )
 \, + \,
 \rho_{i,j} \, \mbox{Tr} (\Delta_i \Delta_i ) 
\mbox{Tr} (\Delta_j^\dagger \Delta_j^\dagger ) ,
\nn \eea

\bea
{\mathbold V}_{\phi \Delta} & = &
\alpha_{i,j,k} \, \mbox{Tr} (\phi^\dagger_i \phi_j ) 
\mbox{Tr} (\Delta_k \Delta_k^\dagger )  \,
+ \, \beta_{i,j,k,l}  \, \mbox{Tr}(\phi^\dagger_j \Delta_k  \phi_i 
\Delta_l^\dagger ) .
\nn \eea
Here we have introduced a shorthand notation in which every term in 
the last equations  stands for the generic term of its type. The full
potential contains all  possible independent combinations 
of the fields of such type and can be found in Ref.\cite{gabriela}.

The presence of the $\beta$ terms in addition to the complexity of the vevs is 
going to bring the desired spontaneous $CP$-violation. 
In fact, it is due to the  $\beta$
terms that the first derivative  equations are no longer homogeneous 
allowing the phase degrees of freedom to survive. 
Previous works had
eliminated these non-diagonal quartic couplings between the two scalar
triplets and the bidoublet in order to avoid  the 
ocurrence  of FCNC at the  minimum of the potential. 
However, as was shown in Ref.\cite{gabriela}, 
these FCNC can be
kept under control and still retain the terms allowing for spontaneous
breakdown of $CP$. 
It is important to notice that, in order to have spontaneous $CP$-violation
in the left-right symmetric model, one needs to have complex vevs in
both  bidoublet and one triplet i.e., one needs two phases.
However, the leptonic sector does not concern us here and the consequences
of $CP$-violation in the quark sector arise only from the Higgs field
$\phi$ belonging to the $(\frac{1}{2},\frac{1}{2},0)$ representation.
Therefore, only the phase $\alpha$ is going to be relevant.

Besides that, 
the minimal left-right symmetric models with spontaneous $CP$-violation
possess the useful property that all the $CP$-violating observables can
be expressed in terms of a single phase, a ratio of scalar vevs, quark
masses and weak mixing angles, but not on unconstrained quantities such
as Yukawa couplings or additional phases. This happens only when the
discrete left-right symmetry and 
$CP$ symmetry are imposed on the Lagrangian as occurs
in our model.

\section{Parametrization of the KM matrix}

An important prelude to the phenomenological study of $CP$-violation
in any model is to identify the number of genuine $CP$ phases. By the
geniune $CP$ phases  we
mean the phases left over when we have used  all our 
freedom to redefine  the particle fields. To carry out 
this procedure, we need to know the structure of the mass matrices
in the model. We, therefore, start with the Yukawa Lagrangian. 

The most general Yukawa Lagrangian for quarks in the left-right model is
given by
\bea
\cal{L}_Y &=& f \bar{\Psi}_L \phi  \Psi_R \; + 
h \bar{\Psi}_L \tilde{\phi}  \Psi_R \; +  \; h.c. ,
\label{yuk}
\eea
where $f$ and $h$ are the Yukawa couplings and the summation over families
is understood.  
A direct consequence of imposing $CP$ as spontaneously broken symmetry,
togheter with the discrete left-right symmetry, is that the Yukawa
couplings matrices $f$ and $h$ in Eq.(\ref{yuk}) must be real and symmetric.
After the SSB the quark mass matrices generated
by $\langle \phi \rangle_0$ are
\bea
M^u & = & 
\frac{1}{\sqrt{2}}\left(f k_1 + h k_2 e^{-i\alpha}\right) =\frac{k_1}{\sqrt{2}}
\left( f + h r e^{-i\alpha} \right), 
\nn \\
M^d & = & 
\frac{1}{\sqrt{2}}\left(h k_1 + f k_2 e^{i\alpha}\right) =\frac{k_1}{\sqrt{2}} 
\left( h + f  r e^{i\alpha} \right) ,
\label{one}
\eea
where $M^u$ ($M^d$) is the up (down) type quark mass matrix and
$r \equiv |k_2|/k_1$.
The only complex parameter in Eq.(\ref{one}) is the complex phase in
$k_2=|k_2| e^{i \alpha}$ which is the unique source of $CP$-violation 
in the charged fermion mass matrices that appear in our model.

Since $M^u$ and $M^d$ are symmetric complex matrices, they can be  diagonalized
 by the orthogonal transformations
\bea
V^u M^u V^{u\,T} =D^u , \nn \\
V^d M^d V^{d\,T} =D^d ,
\label{four}
\eea
where $V^u$ and $V^d$ are unitary matrices and $D^u$ and
$D^d$ the diagonal quark mass matrices.
Since the charged current interaction of the 
theory can be written in the quark mass eigenstate basis as
\bea
\cal{L}_{CC}&=& \frac{g}{\sqrt{2}} \left( W^{\dagger \mu}_L  \,
\overline{u}_L \, K_L \, \gamma_\mu \, d_L \; + \; 
W^{\dagger \mu}_R \,
\overline{u}_R \, K_R  \, \gamma_\mu \, d_R \right) \; + \; h.c. ,
\label{char}
\eea
where $K_L$ and $K_R$ are the left and right KM matices, 
it follows 
 that by choosing to diagonalize the quark mass matrices in the 
form (\ref{four})
we have implicitly fixed our phase convention for quarks in such a way 
that the relation between $K_L$ and $K_R$ is
\bea
K_L = V^{u \,\dagger} V^{d } = K_R^*.
\label{three}
\eea
Therefore, the KM angles in $K_L$ and $K_R$ are equal and the total
number of independent phases in both matrices together  
is  the same as one unitary matrix can contain,
which is $\frac{1}{2}N(N+1)$.  Performing an appropriate rephasing 
of the quark fields 
some of the phases can be shifted from the left sector to 
the right one and vice versa, but, in general, not all of them can be
removed from one matrix to the other.

For a moment we will work in the basis
(\ref{three}). However, for our phenomenological 
analysis in three generations it is more convenient to choose a basis in which
there is only one phase, the KM matrix phase $\delta$ of the SM, 
left in $K_L$. This allows us to use the SM expressions for the
$CP$-violating observables coming from the left sector. 
 In the two generation analysis all the phases can be shifted to
$K_R$.

Eliminating the matrices $f$ and $h$ in Eq.(\ref{one}) we arrive at a
matrix equation of the form 
\bea
(1-r^2) W D^u W + (r^2 e^{2i\alpha} - 1) D^u = 2 i r \sin\alpha \, 
K D^d K^T,
\label{ex}
\eea
where $W=V^{u\dagger} V^{u\ast}$ 
is a unitary symmetric matrix and $K=K_L$ in the 
representation (\ref{three}). This equation is 
exact for any number of generations and cannot be solved exactly. 
In the first  approximation, inspired by the experimental data, 
in which $K$ is taken to be diagonal one can easily show that 
Eq.(\ref{ex}) has solutions only if the following requirement is 
fulfilled
\bea
\frac{\mid r \sin\alpha \mid}{1 - r^2} \le \frac{ m_b}{m_t}.
\label{con}
\eea
 Putting this into  
another way, in order to give   quarks the experimentally
observed masses through the Lagrangian  (\ref{yuk}) the parameters in
Eq.(\ref{one}) must satisfy the condition (\ref{con}). It has been shown 
\cite{frere} that for three generations already this first approximation 
gives very good results if compared with the 
complete numerical calculation.
   
As will be argued later,  avoiding fine tuning of
  $\alpha$ to extremely small values the expression
(\ref{con}) implies $\mid r \mid \le \cal{O}(m_b/m_t).$ 
Another important consequence of the condition (\ref{con}) is that 
 two independent parameters $r$ and $\alpha$  can be reduced to a
significant one which satisfies
$\mid r \sin\alpha \mid \le m_b/m_t.$ As we will see later, in 
first approximation all the phases 
in $K$ will appear as linear in $r \sin\alpha.$  

Now we  calculate explicitly the phases in the KM matrices in terms of
$r$ and $\alpha.$ 
When $r=0$, the mass matrices have the form
\bea
M^u_0 =\frac{ k_1 f}{\sqrt{2}}, \;\;\; 
M^d_0 =\frac{ k_1 h}{\sqrt{2}},
\eea
 and Eq.(\ref{one}) can be rewritten as
\bea
M^u & = & M^u_0 + r M^d_0 e^{-i\alpha} , \nn \\
M^d & = & M^d_0 + r M^u_0 e^{i\alpha} .
\label{two}
\eea
As suggested by (\ref{con}), we will work under the assumption that
$r$ is so small that we can use it as a small perturbation 
parameter  and 
include only terms of lowest order in $r$.
Under this assumption we can treat the second term in Eq.(\ref{two})
as a small perturbation and solve the unitary matrices which
diagonalize $M^u$ and $M^d$ to the lowest nontrivial order in $r$. The zeroth 
order mass matrices can be parametrized in terms of the quark
masses and mixing angles. By proper choice of the flavour
basis we can assume that $M^u_0$ is diagonal without loss of
generality.  In the same basis, $M^d_0$ can be written in terms of 
the down type quark masses and mixing angles. Of course, the 
quark masses and the angles will be modified when $r$ is 
included, but since we are calculating the complex phases
in the diagonalizing matrices $V^{u,d}$  to the lowest
order in $r$, the corrections to these masses and angles
are almost negligible.

First we will work  assuming only two generations.  The generalization 
to the three generation case will be done afterwards. 
The reason for such an  approach  will become clear a bit later.  
The most general parametrization of $K_L$ and $K_R$ in two generations
can be written as follows
\bea
K_L &=& e^{-i \frac{\gamma}{2}} 
\pmatrix { e^{i \frac{\delta_2}{2}} \cos\theta & 
e^{i \frac{\delta_1}{2}} \sin\theta \cr
- e^{-i \frac{\delta_1}{2}} \sin\theta & 
e^{-i\frac{\delta_2}{2}} \cos\theta \cr} ,
\nn \\
K_R &=& 
e^{i \frac{\gamma}{2}} \pmatrix { e^{-i \frac{\delta_2}{2}} \cos\theta & 
e^{-i \frac{\delta_1}{2}} \sin\theta \cr
- e^{i \frac{\delta_1}{2}} \sin\theta & 
e^{i\frac{\delta_2}{2}} \cos\theta \cr} ,
\label{five}
\eea
where $\theta$ is the Cabbibo angle.
After solving Eq.(\ref{four})
for the complex phases in $V^u$ and $V^d$ 
we obtain the following equations for $\delta_1$, $ \delta_2$ and
$\gamma$ to lowest order in $r$

\bea
\delta_1 &=& r \sin\alpha \left[ \frac{1}{2} \left( 
\frac{A}{m_u} - \frac{B}{m_c} - \frac{C}{m_d} + \frac{D}{m_s} 
\right) + 2 \left( \frac{m_s - m_d}{m_u+m_c} - \frac{m_c - m_u}
{m_s + m_d} \right) \cos^2\theta \right], \nn \\
\delta_2 &=& r \sin\alpha \left[ \frac{1}{2} \left( 
\frac{A}{m_u} - \frac{B}{m_c} + \frac{C}{m_d} - \frac{D}{m_s} 
\right) - 2 \left( \frac{m_s - m_d}{m_u+m_c} - \frac{m_c - m_u}
{m_s + m_d} \right) \sin^2\theta \right], \nn \\
\gamma &=& r \sin\alpha \;\frac{1}{4} \left( 
\frac{A}{m_u} + \frac{B}{m_c} + \frac{C}{m_d} + \frac{D}{m_s} 
\right) ,
\label{signs}
\eea
where
\bea
A &=& m_d \cos^2\theta + m_s \sin^2\theta , \nn \\
B &=& m_d \sin^2\theta + m_s \cos^2\theta , \nn \\
C &=& m_u \cos^2\theta + m_c \sin^2\theta , \nn \\
D &=& m_u \sin^2\theta + m_c \cos^2\theta .
\eea
In these expressions all the quark masses  can be both positive or negative.
Strictly speaking, they are the physical masses (defined to be positive)
with additional plus or minus signs which arise from the Yukawa couplings.
We prefer to keep the signs in the masses instead of absorbing 
them to the phases of the KM matrices. 
Unlike in the SM, where  the observables do not depend on these 
signs of  masses, in the left-right model the signs  
themselves are
observables. Because of this, it is important to keep track of
the signs  and to disentangle their physical significance.
Just by inspection, we can see from Eq.(\ref{signs})
that for a given value of $r \sin \alpha$,
there are as many distinct solutions as there are signs of masses, up 
to an overall sign which is not observable. That is 
$2^5 = 32$ solutions. However, we will show later that 
some of them can be ruled out
on a phenomenological basis. The remaining ones can be 
divided  into two groups depending on the relative  sign of the
SM and  left-right contributions to $\epsilon.$

As we  see, in this model all the $CP$-violating phases in the
hadronic sector can be directly related to $r \sin\alpha.$
This feature is independent of how many generations of quarks we 
have in the model. By choosing  the relative phases between 
quarks fields, in the $N=2$ case, all the phases in $K_L$ can
be removed into $K_R$. However, in the $N=3$ case one phase
will be left over in $K_L$. This is nothing but the well known
observation by Kobayashi and Maskawa \cite{KM}. 
In the following we will work in the basis where the maximum
number of  phases are
shifted to the right sector.

To extend  our model to the
three generation case we have to analyse the effect of the KM 
phase. 
It is well known that the dominat left-right contributions in the kaon system
do not involve the third generation \cite{beall}, and consequently, only the 
phases that are present in the  two generation  $K_R$ 
will be needed in our analysis. Therefore, the choice of $K_R$ in the
form similar to Eq.(\ref{five}) (all the phases should be multiplied by two
due to the shifting them from $K_L$ to $K_R$) is the most general in our case.
To make the phenomenological estimation complete, we will have to use
the complete three generation $K_L$ matrix put in the usual SM form.
In a suitable convention $K_L$ can be written
in the  form
\bea
K_L = \pmatrix { c_1 & -s_1 c_3 & -s_1 s_3 \cr
s_1 c_2 & c_1c_2c_3 - s_2s_3e^{i\delta} & 
 c_1c_2s_3 + s_2c_3e^{i\delta} \cr
s_1s_2&  c_1s_2c_3 - c_2s_3e^{i\delta} &
 c_1s_2s_3 - c_2c_3e^{i\delta} \cr},
\label{kmmat}
\eea
where $s_i \equiv \sin\theta_i$, $c_i \equiv \cos\theta_i$
and $\theta_1$, $\theta_2$ and $\theta_3$ are the three KM
angles.
A lesson we have learned from the Kobayashi-Maskawa 
$SU(2)\times U(1)$ model is that all the $CP$-violating quantities
in the model are proportional to a single  factor $s_1^2s_2s_3 \sin\delta$
which is of the order $\lambda^6$ in the Wolfenstein parametrization.
The dominant left-right contribution to $CP$-violating observables, however,
is of the order of $\lambda^2$ since $CP$-violation in the model can occur
with two generations only. Even if the right-handed
charged currents are suppressed by the large $M_2$ and have a very
little effect in low energy $CP$-conserving quantities (we assume this
to be always true and use the SM values for the KM matrix entries)
in $CP$-violating observables the left-right  part can possibly 
dominate over the 
SM one which is strongly suppressed. Therefore the kaon system is very 
good to search for the left-right symmetry. 

Consequently, if in our model with spontanous $CP$ breaking $\delta$
itself is of the same order of magnitude as $\delta_1$ or
$\delta_2$ (which we have calculated before), then we
expect the SM contribution to be somewhat suppressed.
To show that,  we have to compute the relation between 
$\delta$ and $r \sin\alpha$. Following the
same procedure as in the four quark case, after tedious algebra we
obtain
\bea
\delta = r \sin\alpha \, \frac{m_c}{m_s} \left(\frac{s_2+s_3}{s_3}
\right) \left[ 1 + s_3\left(s_2 +s_3 \right) \frac{m_t}{m_c}
\right] .
\label{smdel}
\eea
With this result we can proceed to the phenomenological 
analysis of our model of spontaneous $CP$-violation.

\section{ Constraints on the left-right model parameters from 
$\epsilon$ and $\epsilon'/\epsilon$}

There exists already an extensive literature on  $|\Delta S|=1,2$
effective interactions of  kaon system in the left-right symmetric models.
Since our aim is to perform a 
phenomenological analysis of our model of spontaneous $CP$-violation 
we will adopt the already known 
expressions of $\epsilon$ and $\epsilon'$
together with the estimations of hadronic matrix
elements, QCD short distance corrections and final state interactions 
 from the most comprehensive 
works \cite{frere,bur}. We note here that for the hadronic
matrix elements of the right-handed sector
the vacuum saturation approximation is used which is assumed to
give precise enough results for our analysis.
We will update the previous expressions by using the recent experimental data
for quark masses as well as for the modulus of the  KM matrix elements. 
After substituting the KM matrix elements from the previous Section
we can use the resulting formulae
to constrain the model parameters.

In the left-right model the effective $|\Delta S|=2$ Hamiltonian gets 
contributions from the box diagrams presented in Fig.1. There are two charged 
current gauge boson mass eigenstates $W_1$,  predominantly 
the left-handed $W_L$,  and $W_2$,  predominantly the right-handed $W_R$,
in the model. Their mixing angle is very small and can be expressed as
\bea
\zeta=\frac{2 r}{1+r^2}\left( \frac{M_1}{M_2}\right)^2,  
\eea
where $M_1$ and $M_2$ denote  masses of the  corresponding gauge bosons.
The diagram   with both $W$-s being the ordinary left handed 
$W_1$ gives  the SM contribution. 
However, in our model the phase $\delta$ is not an independent 
parameter but related to $\alpha$ by Eq.(\ref{smdel}).
The diagram with two
$W_2$-s is negligible compared with the SM one because of  very 
large mass
of the new gauge boson. The couplings of charged Higgs bosons with quarks
 are suppressed by the factors of $m_q/M_{H}$ and 
since their masses are of the order of the right-handed breaking scale
one can ignore them.
  Therefore, the dominant left-right 
contribution to the effective Hamiltonian of $|\Delta S|=2$ comes from the
diagrams with one $W_1$ and one $W_2$. Since the contribution coming from
these diagrams is proportional to the mass squared of
the up-type quarks which run in the loop,  one may expect 
that diagrams with top quarks are dominant. However, due to the small
off-diagonal elements of the KM matrix for the third generation, the 
contribution from the top quark diagrams relative to the c quark diagrams
 is of the order  $\lambda^8 m^2_t/m_c^2 \sim 10^{-2}$, 
where $\lambda\sim 0.2$
is the Wolfenstein's expansion parameter.  
This proves that within 
our assumption of the $CP$ symmetry of Lagrangian
(\ref{yuk}), which makes the modulus of the left and right KM matrices to be
equal, we can safely neglect the third generation in dealing with 
the left-right contribution to the parameter $\epsilon.$

Let us now consider the parameter $\epsilon=\epsilon_{SM}
+\epsilon_{LR}, $ where $\epsilon_{SM}$ comes from the left-left 
and $\epsilon_{LR}$ from the left-right box diagram. 
The expressions we are dealing with can be written as  \cite{frere,bur} 
\bea
\epsilon_{SM}=e^{i \pi/4}\; 1.34 \; s_2 s_3 \sin\delta \left[1+860\; 
S\left(\frac{m_t^2}{M_1^2}\right) s_2 Re V_{ts} \right],
\label{esm} 
\eea
where 
\bea
S(x)=x\left[ \frac{1}{4}+\frac{9}{4}\frac{1}{(1-x)}-
\frac{3}{2}\frac{1}{(1-x)^2} \right] -
\frac{3}{2} \left[ \frac{x}{1-x} \right]^3 \ln x
\eea
and 
\bea
\epsilon_{LR}=-e^{i \pi/4}\; 0.36 \sin(\delta_2-\delta_1)
\left[ \frac{1.4\; TeV}{M_2} \right]^2 \left( 1+0.05\ln
\left[  \frac{M_2}{1.4\; TeV}\right]\right).
\eea
Here $\delta,$ $\delta_1$ and $\delta_2$ are the KM phases calculated
in Section 3 and $s_2$ and $s_3$ are defined in Eq.(\ref{kmmat}).
For our computations we have used the  numerical values \cite{pdb}
$s_1=0.2209\pm 0.0027, $
$s_2=0.0430\pm 0.0258$ and $ s_3=0.0158\pm 0.007,$ which have been
obtained in the SM. We assume that the effect of the new heavy scale
in determining the KM matrix elements 
can be neglected at the first approximation.
 Since in the expressions of the $CP$-violating phases the quark masses 
 appear  as  ratios then the result does not depend
on the mass scale at which they are taken. We can therefore choose the running 
masses  at $Z_0$ scale 
\cite{qmass}: $m_u=(1.5\pm 1.2) $ MeV, $m_d=(4.1\pm 1.7) $ MeV, 
$m_s=(83.\pm 30.) $ MeV, 
$m_c=(0.52\pm 0.10)$ GeV and $m_t=(180.\pm 13.)$ GeV.

The $CP$-violating parameter $\epsilon=|\epsilon| e^{i\phi}$ 
has been measured with a good
accuracy  $|\epsilon_{exp}| =(2.26 \pm 0.02) \cdot 10^{-3}$ 
and $\phi\approx \pi/4$ \cite{pdb}. 
We fix it to the 
experimentally measured value and vary the free parameters of the model,
$M_2$, $r$ and $\alpha$ as well as the signs of the quark masses, 
in such a way as to get the correct $\epsilon.$

Let us study first how the changes of the signs  of quark masses affect
the value of $\epsilon$ (for a moment we leave the phase $e^{i\pi/4}$ aside).
In the case of all positive masses $\epsilon_{SM} > 0 $ and 
$\epsilon_{LR} < 0$. Changing the sign of $m_d$ changes the sign of
$\epsilon_{LR}.$
It is easy to check that  changing   the sign of $m_s$ changes the sign of 
$\epsilon_{SM},$ while changes in  signs of the other quarks either
cange the signs of the both contributions ($m_c$) or leave them 
unchanged ($m_t$). Since  $\epsilon_{SM}+\epsilon_{LR}$ should be positive
then the situation where both left-left and left-right contributions
are negative cannot be realized. Therefore, just on this basis, we can
exclude some of the  combinations of quark masses as  solutions.
In principle, we have two qualitatively different situations: 
either one of the contributions 
$\epsilon_{SM}, $  $\epsilon_{LR }$ is negative and another positive or
both of them are positive.
Therefore, different models can be classified according to the relative sign
of the SM and LR contributions to $\epsilon$: Class I if they are different
and Class II if they are the same.  In general, this classification can
be done also by the relative sign of $m_d$ and $m_s$ but one has to
remember that, on the contrary to claims in Ref.\cite{eg,frere},
 not all combinations of the signs are viable solutions.

In Fig. 2. we plot $M_2$ against $r \sin\alpha$ for two different
choices of  signs of the quark masses using the central values of  all
the experimentally measured input parameters. For the curve $(a)$ $m_d$
and $m_s$  are taken to be negative and the rest of the masses
positive while for the
curve $(b)$ only $m_d$ is taken  to be negative. 
These two choices belong to 
 the two distinct classes of models,
Class I and II, respectively.  We have checked that
choosing different combinations 
of the signs of  quark masses inside the classes the curves in Fig.2  change
not more than  $\sim 20\%$.
To present all of them will not enlight the disscusion at all. Therefore,
we plot just one representative 
curve from each class.
  
Our complete analysis shows that $r \sin\alpha$ is limited to the region
\bea
0.0005 \le r\sin\alpha \le 0.017,
\label{x}
\eea
where the upper limit comes from Eq.(\ref{con}).
The use of smaller values of $r\sin\alpha$ 
would lead to lighter $W_2$ than tolerated by 
the lower limit coming from the $K_S$-$K_L$ mass difference \cite{frere,beall}
if one wants to get the correct $\epsilon_{exp}.$    

With $r \sin\alpha$ in such a range  and with the central values of
experimental data  the SM contribution alone is
always smaller than $|\epsilon_{exp}|$ and we need the LR part to 
agree with  the experiment.  
Therefore, also these combinations of signs of quark masses for which
$\epsilon_{LR} < 0$ are not allowed in the present case.
In the case of Class I (Fig.2 $(a)$),  $r \sin\alpha$ can vary
over all the  values of (\ref{x}) and $M_2$ is a slightly increasing 
function of $r \sin\alpha$  with the maximum value $M_2\approx 5.5$ TeV.   
However, the behaviour of Class II (Fig.2 $(b)$) is  completely  different.
In this case $r \sin\alpha$ has a lower limit around $0.0045$ and
$M_2$ increases very fast when $r \sin\alpha$ approaches the maximum value.
 This behaviour can be easily understood. Since in Class II the SM and LR
contributions have the same sign and the SM part is getting bigger 
 if $r \sin\alpha$ is increasing then we need just a small additional
contribution from the LR sector. Therefore, $M_2$ should be larger.
As suggested by  grand unified 
theories \cite{gut} the natural values for $r$ are
around $10^{-3}.$ In light of this prediction Fig.2 clearly
 prefers very light $M_2$ and models belonging to Class I. 

In order to see the allowed space for $r$ and $\alpha,$ in Fig.3 we plot 
  $r$ against $\alpha$ for fixed $r \sin\alpha=0.001.$
For most of $\alpha$ values $r$ is quite flat and our previous
discussion is, indeed, valid for most of the parameter space.
 $r$  increases fast
only for very small or close to $\pi$ phases. 
For  models of spontaneous $CP$-violation these extreme values of 
$\alpha$ are unnatural since the vevs of the bidoublet 
are almost real without any deeper reason. However, there is a more
strict argument to prohibit very small values of $\sin\alpha.$ 
In order to provide quarks with the experimentally measured masses
and keep $\alpha$ to be very small at the same time we have to
make some of the Yukawa couplings in the Lagrangian (\ref{yuk}) to be
large and the present  perturbative calculation is not valid any more.
Therefore, in the framework of  
the perturbation  we have performed, the extreme values of $\alpha$ 
are not allowed. We will assume this in the following.

So far, we have not taken into account the effects of the
 experimental errors.
Looking at the numerical values of the quark masses and $s_2$ and $s_3$
we see that the least accurately determined parameters are $m_s$ and
$s_2.$ Indeed, a  numerical analysis shows that our results are most sensitive
to the changes of $s_2$ which gives the dominant error. 
If we tune the input parameters within 
what the experimental results can tolerate
in such a
way that the SM contribution to $\epsilon$ can be bigger than 
$|\epsilon_{exp}|$ then we have a qualitatively new situation which should be
analysed.

In Fig.4 we plot $M_2$ against $r \sin\alpha$ for the same class of
models as in Fig.2 but taking for $s_2=0.0688$  i.e. the extreme
value allowed by the 1 standard deviation experimental error ( $ 
68 \%$ C.L ).
With this $s_2,$  $|\epsilon_{SM}|$ is almost  maximized by the experimental
errors since they are dominated by $s_2.$ The curve $(a)$ in Fig.4 
corresponds to the curve $(a)$ in Fig.2. However, for the Class I model
we have now another curve $(b)$ which also gives the correct $\epsilon.$
Since  $|\epsilon_{SM}|$ can be bigger than $|\epsilon_{exp}|$ we have two
possibilities: either $\epsilon_{LR}$ is large and $\epsilon_{SM}$ 
reduces it to the correct   $|\epsilon_{exp}|$ value (curve $(a),$ 
$m_d, \; m_s <0$)
 or $\epsilon_{LR}$ is small and serves to reduce $\epsilon_{SM}$ 
to the needed value (curve $(b),$ $m_d, \; m_s >0$). 
The curve $(c)$ in Fig.4 denotes
the behaviour of $M_2$ in the case of Class II model.

As can be seen from the asymptotic behaviour of curves $(b)$ and $(c),$
there is a pole in $M_2$ corresponding to  a $r \sin\alpha$ 
value for which
the SM contribution gives exactly the measured $|\epsilon_{exp}|.$
Obviously, for such a $r \sin\alpha$ curves $(b)$ and $(c)$ should go to
infinity.  Assuming the GUT suggested value for $r$ of $10^{-3},$
$M_2$ must have mass around 2-3 TeV. Since the curve $(c)$ can extend only up
to the pole,  the parameter space in the case of Class II models
is even more restricted  than previously 
which does not favor these models. For large
$r \sin\alpha$ there are two values of $M_2$ possible in Class I
models but for the most of the $r \sin\alpha$ space $M_2$ should be
in the range 4-10 TeV. 
If we want to make $M_2$ to be  heavy, say 
heavier than 20 TeV,
we have to do the following. We have to fix the KM matrix entries and
quark masses in a way to ensure $|\epsilon_{SM}| \ge |\epsilon_{exp}|$
at least for some  $r\sin\alpha.$ Then we have to fine tune the parameters
$r$ and $\alpha$  to the very small region where the relation 
$\epsilon_{SM} \approx  \epsilon_{exp}$ 
holds almost exactly.
 And even doing so we still have 
a possibility to explain $\epsilon_{exp}$ by the curve $(a)$ in Fig.4.
We have to conclude that the left-right model with spontaneous 
$CP$-violation clearly prefers light $M_2.$

For completeness we will now analyse $\epsilon'/\epsilon$ in our model.
The effective $\Delta S=1$ Hamiltonian, giving rise to $\epsilon',$
gets contributions from the penguin as well as the tree
level diagrams that are  depicted in Fig.5. The important
left-right contributions come from  $W_2$  exchange and  also from
the $W_1$-$W_2$ mixing. The top quark contribution to the right
sector is small \cite{frere}
 and the two family parametrization should work well.
Again, we adopt formulae for $\epsilon'$ from the previous works
and actualize them by using more precise values for the running $\alpha_s$
and quark masses.
One has
$\epsilon'=\epsilon'_{SM}+\epsilon'_{LR},$
where \cite{frere,bur}
\bea
\epsilon'_{SM}=e^{i(\pi/4+\delta_2-\delta_0)}\; 3.2\cdot 10^{-2} s_2 s_3
\sin\delta\; H(m_t)
\eea
and 
\bea
\epsilon'_{LR} & = & e^{i(\pi/4+\delta_2-\delta_0)} 10^{-2} 
\left\{ \left[
6.8\left[\frac{\alpha_s(\mu)}{\alpha_s(M_2)}\right]^{-2/b} -0.30
\left[\frac{\alpha_s(\mu)}{\alpha_s(M_2)}\right]^{4/b}\right]
\right. \nn \\
& &  \left. \frac{M_1^2}{M_2^2}\sin(\delta_2-\delta_1) 
 +102 \zeta[\sin(\gamma-\delta_1)+\sin(\gamma-\delta_2)] \right. \nn \\
& & \left. -9.6 \zeta [\sin(\gamma+\delta_1)+\sin(\gamma+\delta_2)] \right\},
\eea
where
\bea
\left[\frac{\alpha_s(\mu)}{\alpha_s(M_2)}\right]^{a/b}=
\left[\frac{\alpha_s(\mu)}{\alpha_s(m_b)}\right]^{3 a/25}
\left[\frac{\alpha_s(m_b)}{\alpha_s(m_t)}\right]^{3 a/23}
\left[\frac{\alpha_s(m_t)}{\alpha_s(M_2)}\right]^{a/7}.
\eea
Here $H(m_t)=0.04$, $b=11-2/3 \;n_{flavours}$
 and the phases $\delta_2-\delta_1\approx 40^0$ in the exponential
 are the the strong interaction $\pi\pi$ phase shifts
 (do not mix up with
the phases of the KM matrix).   As the input value for the running 
$\alpha_s$ we use $\alpha_s(M_Z)=0.118$ and evaluate it with 
 the one-loop equation
\bea
\alpha_s^{-1}(m)=\alpha_s^{-1}(M_Z)+\frac{b}{2\pi} \ln(\frac{M_Z}{m}).
\eea

There are two contradicting measurements of $\epsilon'/\epsilon.$
NA31 experiment at CERN claims the result $\mbox{Re}\left(
\epsilon'/\epsilon \right) = (2.3\pm0.7)
\cdot 10^{-3},$ while E731 result at Fermilab is compatible with zero
$\mbox{Re} \left( 
\epsilon'/\epsilon \right) = (0.60\pm0.69) \cdot 10^{-3}.$ Therefore, one can
conclude that $\mbox{Re} \left(
\epsilon'/\epsilon \right)$ should be smaller than a few times
$10^{-3}.$

In Fig.6 we plot $\mbox{Re}\left(
\epsilon'/\epsilon \right)$ of the models of Class I 
($m_d$ and $ m_s$ negative) and II (only $m_d$ negative) 
against $r \sin\alpha$ 
for two different values of $\alpha=\frac{\pi}{2}$ 
and $\alpha=\frac{\pi}{30}.$
Note that for every value of
$r\sin\alpha$ there  corresponds a different $M_2$ which can be
determined from Fig.2.
An interesting result is that in Class 	I models $\mbox{Re}\left(
\epsilon'/\epsilon \right)$
is always negative and in Class II models positive.  
The absolut values of $\epsilon'/\epsilon$
 are in the range of $10^{-5}$ - $10^{-3}$ being notably smaller for
 the maximum $\sin\alpha$ than for  the 
small values of $\sin\alpha.$ 
This is an effect of having larger $r$ in the latter case.
As was argued before,   $\sin\alpha$ cannot be too small since 
we want our expansion in powers of 
$r $ to remain  valid.   In the case of Class II
model, in which $W_2$ should be sufficiently heavy,
 we see the suppression of  $\epsilon'/\epsilon$
due to the large $M_2$ at large $r\sin\alpha.$
 
 $\epsilon'/\epsilon$ is rather sensitive to the change of sign of
 top quark mass, for some parameters it can be modified almost by a
factor of two. In principle, with sufficient accuracy, 
this dependence may shed light on the 
sign in  $\epsilon'/\epsilon$ experiments. 
The dependence of  $\epsilon'/\epsilon$ on
 changes of any input parameters inside the allowed errors is typically of
the order of $\sim 20\%.$ But even with these changes 
our conclusions remain the same.
In fact,  we could not find any allowed region of the parameter space
in which $\epsilon'/\epsilon$ violates the experimental limit.
In general, however, the left-right model with spontaneous $CP$-violation 
seems to prefer  values of $\mbox{Re}\left(
\epsilon'/\epsilon \right)$ around
$ 10^{-4}$ in magnitude.

\section{Conclusions}

Motivated by the possibility of constructing  phenomenologically 
consistent left-right symmetric models with spontaneous breakdown 
of $CP$ symmetry \cite{gabriela} we carry out the analysis of the 
model using $CP$-violating observables in the $K$ system.
We parametrize the general KM phases in terms of a single phase $\alpha$
which comes from the vevs of the bidoublet and is the only source
of $CP$-violation in our model. Due to this fact, and also due to
the  heavy top quark mass which  forces the ratio $r$ of two
bidoublet vevs to be very small, we find the parameter space of the 
model to be rather restricted. We adopt the expressions for 
$\epsilon$ and $\epsilon'$ together with hadronic matrix elements and 
QCD corrections  from Ref.\cite{frere,bur} and
actualize them by updating the values of quark masses, KM matrix elements
and strong coupling constant. 

Using the measurement of $\epsilon_{exp}$ we find that for  most of the
parameter space the mass of the new right-handed gauge boson $W_2$
should be below 10 TeV. Considerably higher masses of $W_2$ can be
achieved only by fine tuning the KM matrix elements, quark masses and
parameters $\alpha$ and $r.$ But even in this case there is another, small
value of $M_2$ which also gives the correct $\epsilon.$
This happens because there are two different classes of modes which can be 
classified according to 
the relative sign of left-left and left-right contributions to 
$\epsilon.$ 
The signs of quark masses
are observables in  left-right models and, unlike in the previous works,
we found that many  combinations of the signs are not allowed by the data.
In the context of grand unified theories which predict the value of $r$
to be $10^{-3}$ \cite{gut} our analysis seems to prefer Class I models in
which $m_d$ and $m_s$ have the same sign.

All the predicted 
values of $\mbox{Re}\left( \epsilon'/\epsilon \right)$ 
for the allowed parameter space
(keeping $\epsilon$ fixed to the experimental value) are below
the accuracy of the present experiments. The most favoured  range
of $\mbox{Re} \left(\epsilon'/\epsilon \right)$  
is around $10^{-4}$ in magnitude being positive for
Class II and negative for Class I models. In the light of
grand unification  for very small $r$ this means that $\mbox{Re}\left(
\epsilon'/\epsilon \right)$
in the left-right models with spontaneous $CP$-violation is negative.

\subsection*{Acknowledgement}

We thank F. Botella, G. Ecker, J. Maalampi, A. Pich, J. Prades
 and A. Santamar\'{\i}a 
for clarifying discussions.
G.B. acknowledges the Spanish Ministry of
Foreign Affairs for a MUTIS fellowship and M.R. thanks the
Spanish Ministry of Science and Education for a postdoctoral
grant at the University of Valencia. This work is supported by CICYT under 
grant AEN-93-0234.

\newpage

{\large\bf Figure captions}

\vspace{0.5cm}

\begin{itemize}

\item[{\bf Fig.1.}] Feynman diagrams contributing to $|\Delta S|=2$ transition.
\\

\item[{\bf Fig.2.}]
$M_2$ as a function of $r\sin\alpha$ for Class I (curve $a,$ $m_d$ and $m_s$
negative)
and Class II (curve $b,$ $m_d$ is negative) models. 
Central values of all
experimental data are used.
\\

\item[{\bf Fig.3.}] 
$r$ as a function of $\alpha$ for fixed $r\sin\alpha=0.001.$
$\alpha$ values close to 0 or $\pi$ are not allowed 
in order to  keep our perturbative results valid.
\\

\item[{\bf Fig.4.}]
The same as in Fig.2 with experimental data tuned to give
the largest  $\epsilon_{SM}.$ Curves $a$ and $b$  correspond to
the possible values of $M_2$ in Class I and curve $c$ in Class II,
respectively.
\\

\item[{\bf Fig.5.}]
Feynman diagrams contributing to $|\Delta S|=1$ transition. 
\\

\item[{\bf Fig.6.}]
$\mbox{Re}\left( 
\epsilon'/\epsilon \right)$ as a function of $r\sin\alpha$ in the case of
two  values of $\alpha.$ The  curves in the positive side
 belong to  Class II
and in the negative side  to Class I.

\end{itemize}

\newpage

\begin{figure*}[hbtp]
\begin{center}
 \mbox{\epsfxsize=12cm\epsfysize=5cm\epsffile{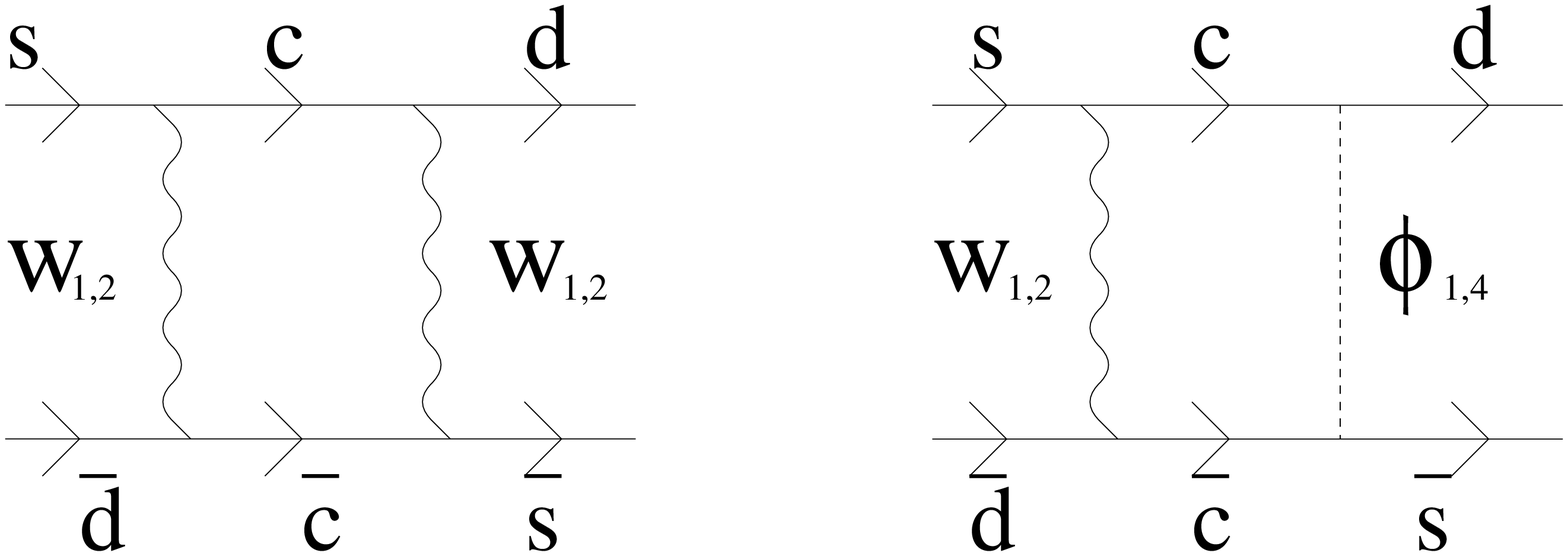}}
\caption{
}
\end{center}
\end{figure*}

\begin{figure*}[hbtp]
\begin{center}
 \mbox{\epsfxsize=14cm\epsfysize=14cm\epsffile{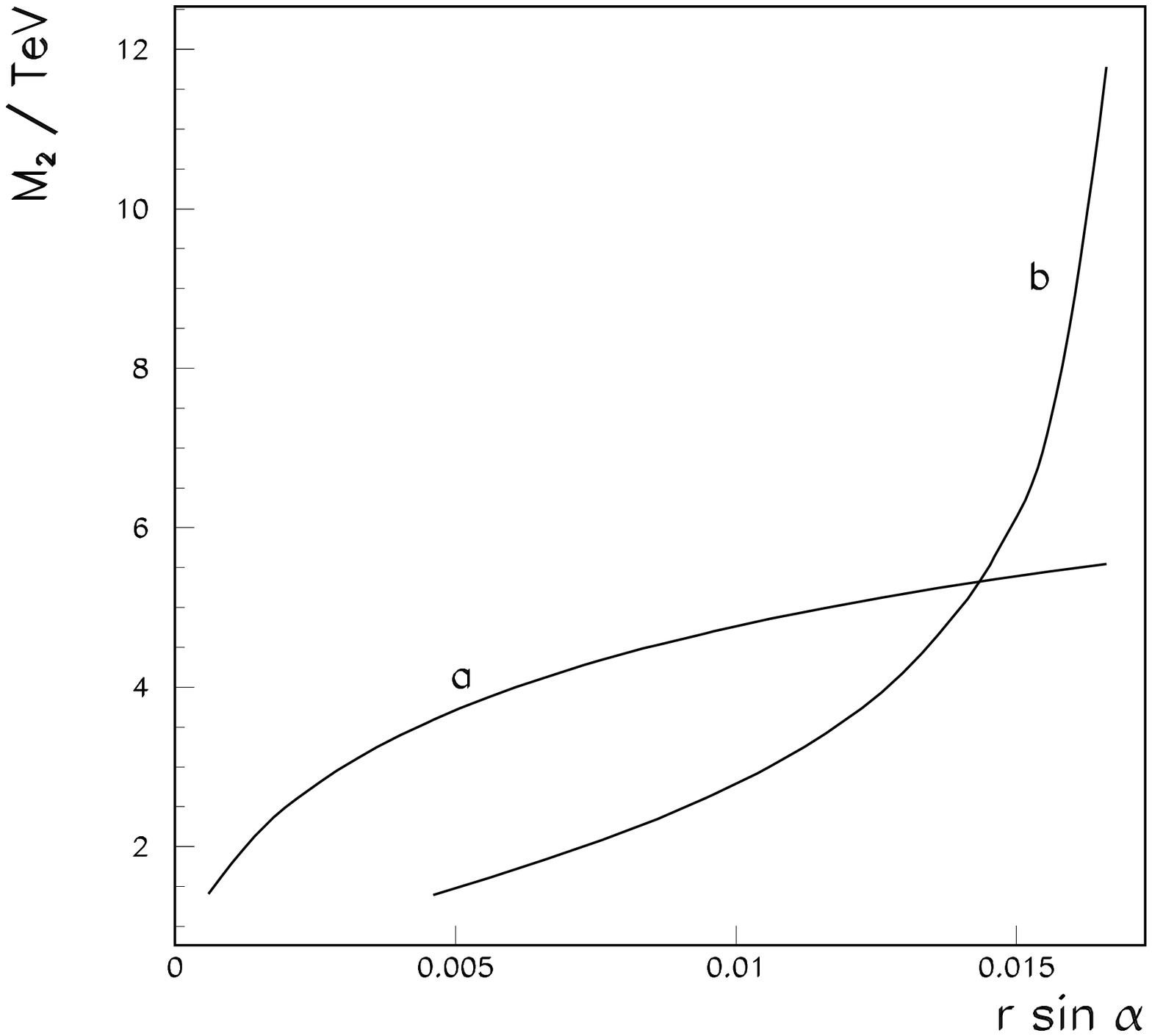}}
\caption{
 }
\end{center}
\end{figure*}

\begin{figure*}[hbtp]
\begin{center}
 \mbox{\epsfxsize=14cm\epsfysize=14cm\epsffile{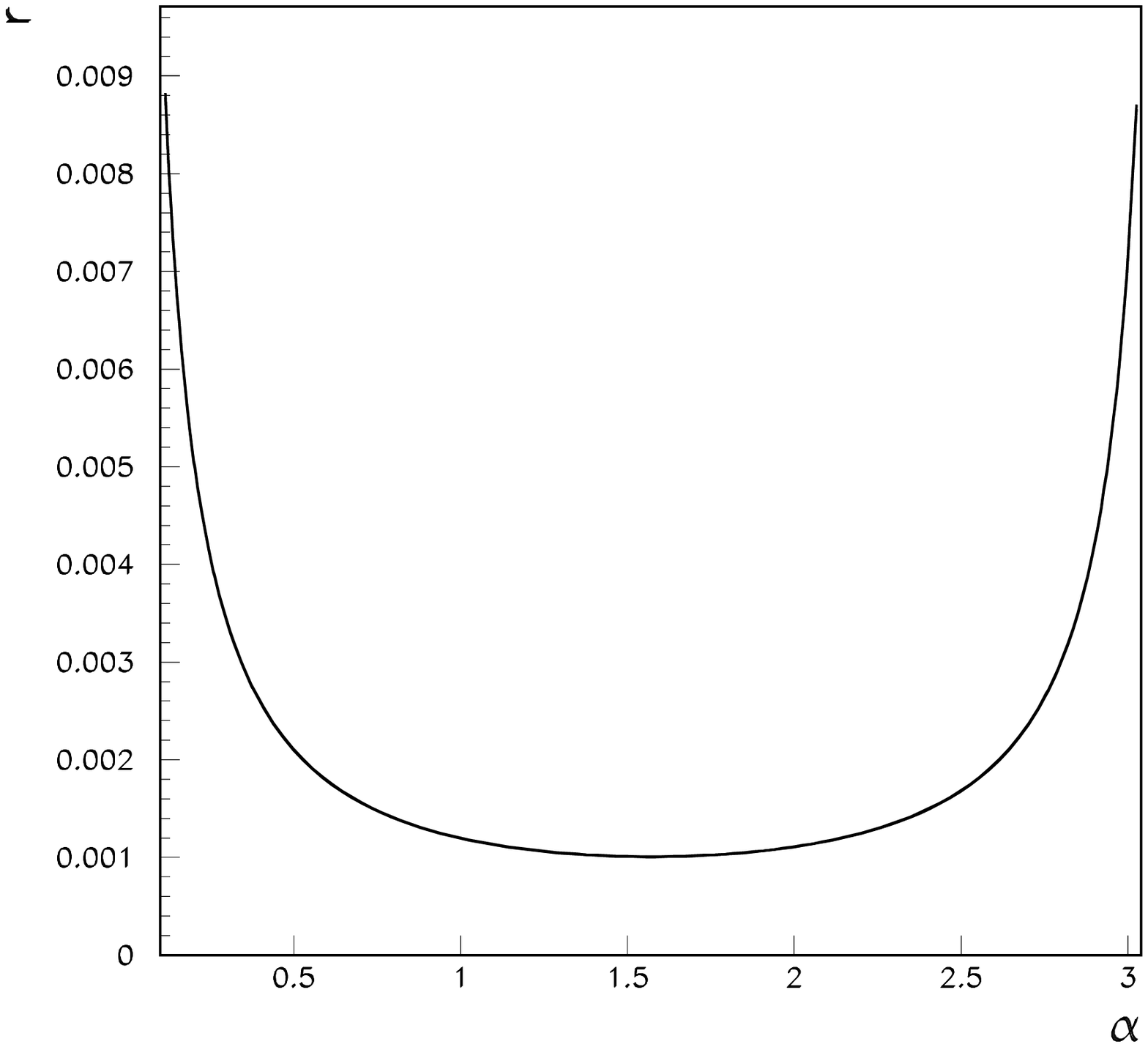}}
\caption{
 }
\end{center}
\end{figure*}

\begin{figure*}[hbtp]
\begin{center}
 \mbox{\epsfxsize=14cm\epsfysize=14cm\epsffile{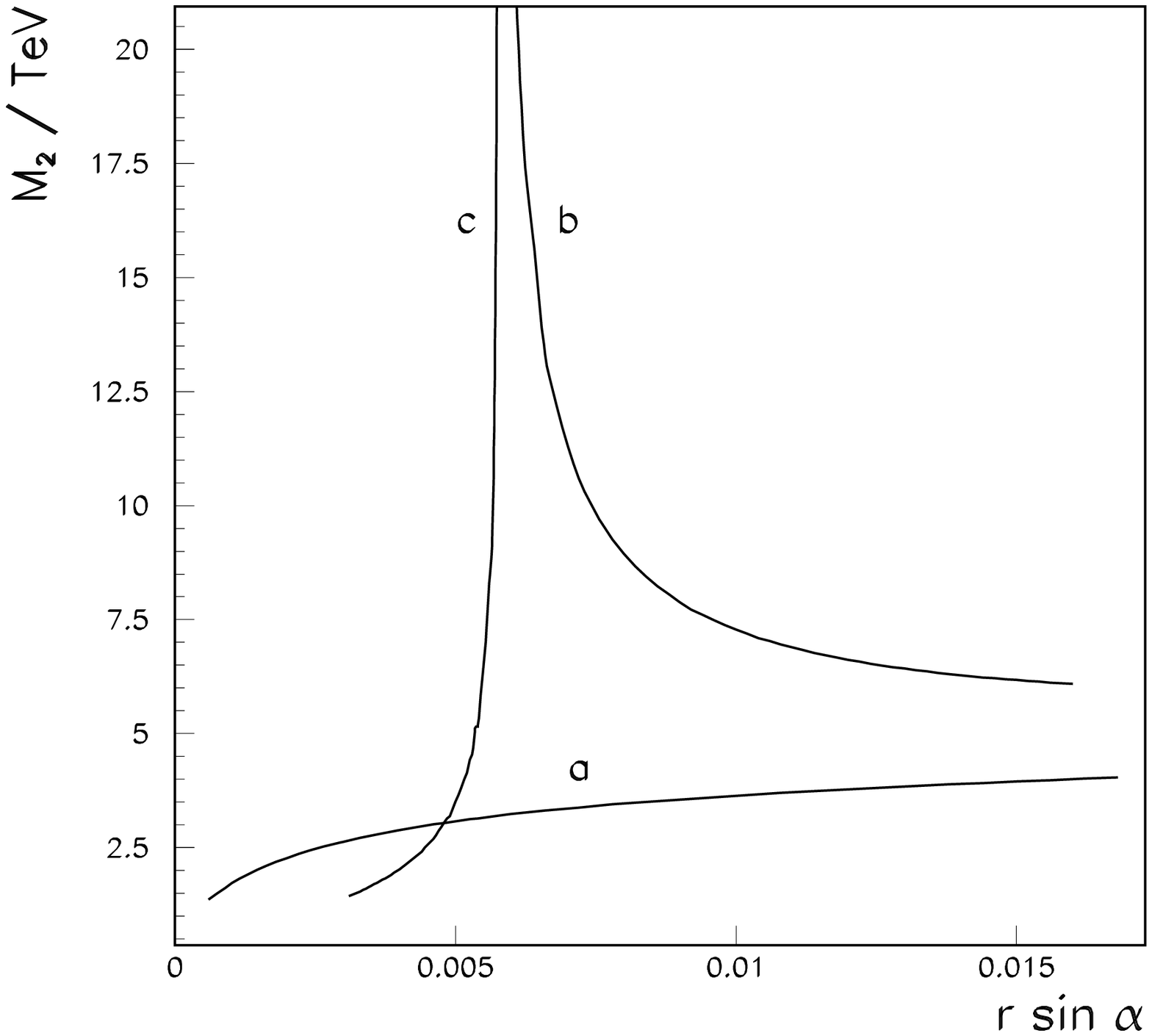}}
\caption{
}
\end{center}
\end{figure*}

\begin{figure*}[hbtp]
\begin{center}
 \mbox{\epsfxsize=12cm\epsfysize=6cm\epsffile{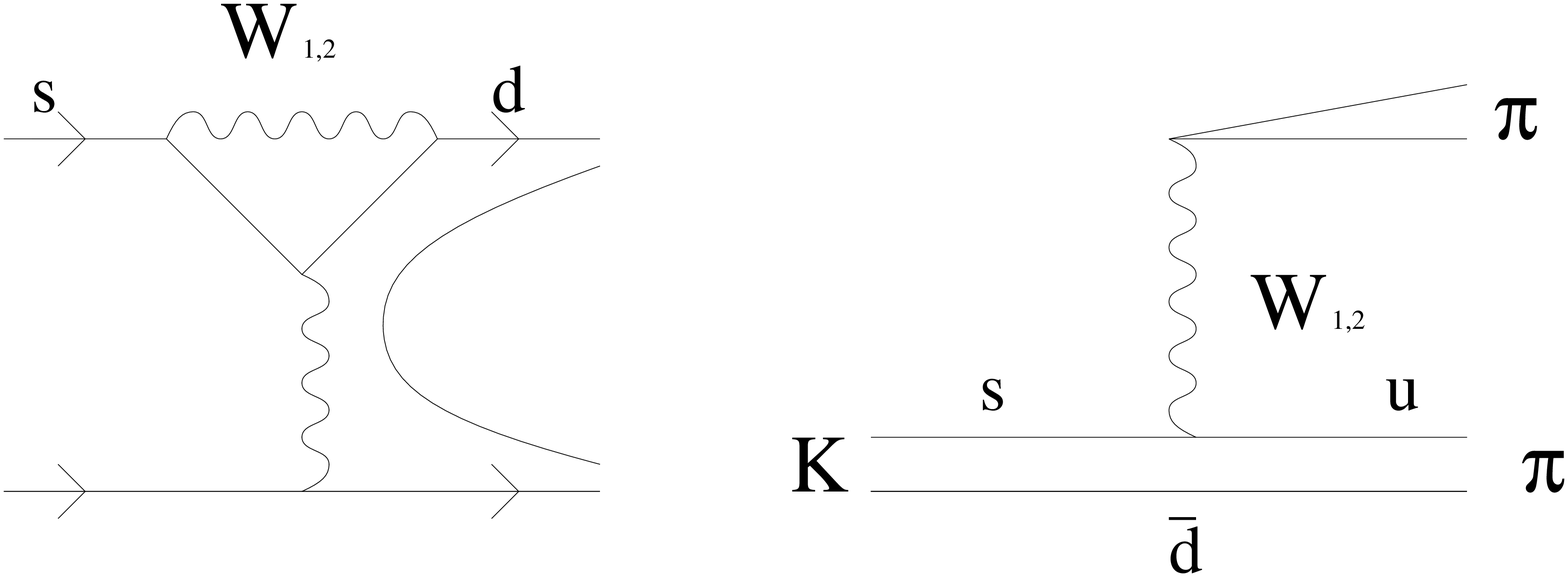}}
\caption{
}
\end{center}
\end{figure*}

\begin{figure*}[hbtp]
\begin{center}
 \mbox{\epsfxsize=14cm\epsfysize=14cm\epsffile{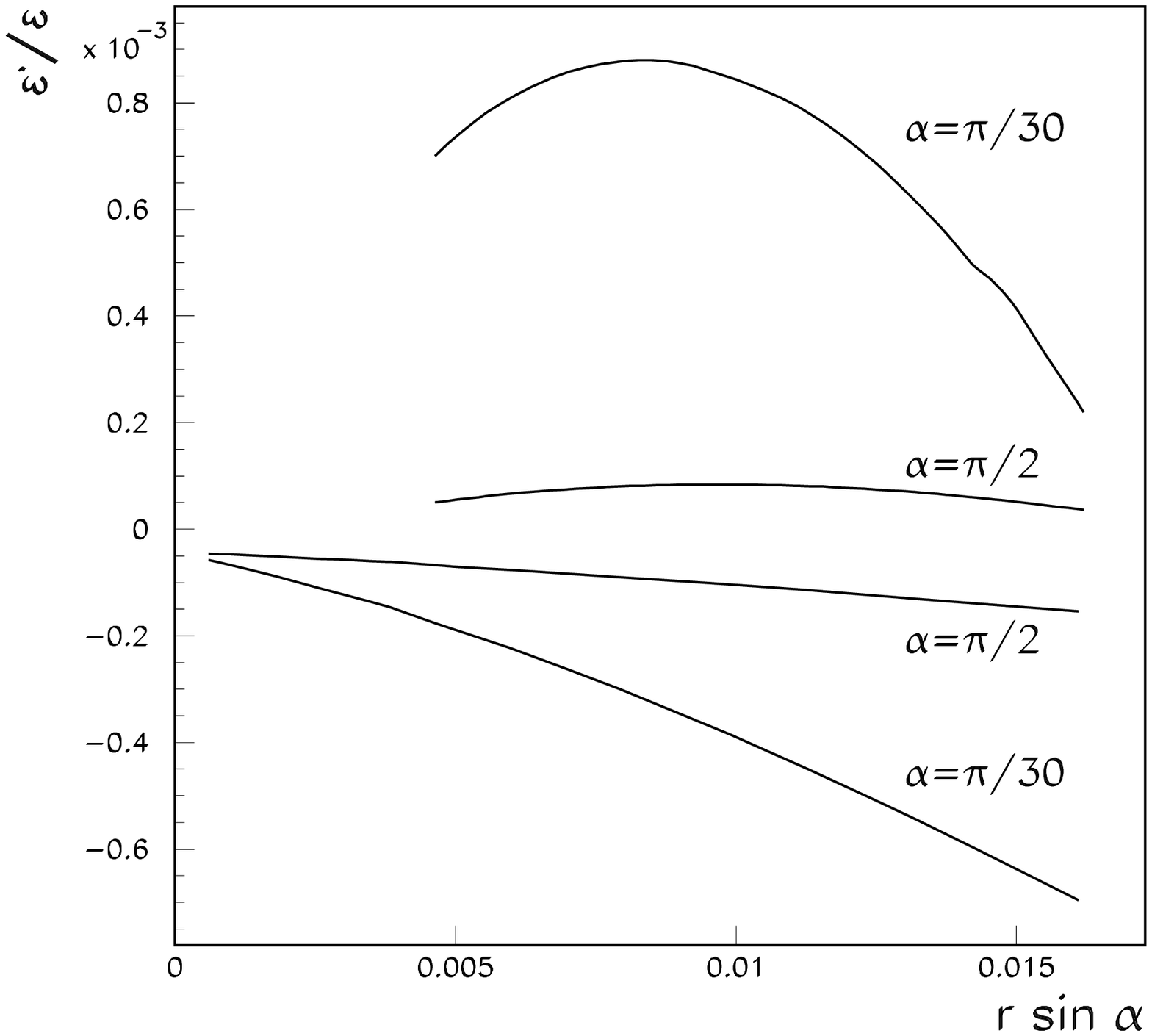}}
\caption{
 }
\end{center}
\end{figure*}


\begin{thebibliography}{99}

\bibitem{jarls1} For a review see e.g. `` CP Violation," ed. C. Jarlskog,
World Scientific (1993).  

\bibitem{KM} M. Kobayashi and T. Maskawa, Prog. Theor. Phys {\bf 49}
(1973) 652.


\bibitem{epsp} F. J. Gilman and M. Wise, \pl{B83} (1979) 83,
\pr{D20} (1979) 2392; \\
B. Guberina and R. Peccei, \np{B163} (1980) 289.

\bibitem{pdb} {\it Review of Particle Properties}, \pr{D50} (1994) 1173.

\bibitem{uni} M. B. Gavela, P. Hernandez, J. Orloff and O. Pene, 
Mod. Phys. Lett. {\bf A9} (1994) 795; \\
M. B. Gavela, P. Hernandez, J. Orloff and O. Pene, 
Nucl. Phys. {\bf B430} (1994) 382.

\bibitem{lr} J. C. Pati and A. Salam, Phys. Rev. {\bf D10} (1975) 275; \\
R. N. Mohapatra and J. C. Pati, Phys. Rev. {\bf D11} (1975) 566 and 2558.


\bibitem{ss} R. N. Mohapatra and G. Senjanovic, Phys. Rev. Lett. {\bf 44} 
(1980) 912, Phys. Rev. {\bf D 23} (1981) 165.

\bibitem{chang} D. Chang, Nucl. Phys. {\bf B 214} (1983) 435.

\bibitem{eg} G. Ecker and W. Grimus, Nucl. Phys. {\bf B258} (1985) 328.


\bibitem{frere} J.-M. Frere, J. Galand, A. Le Yaouanc, L. Oliver, O. Pene
and J.-C. Raynal, Phys. Rev. {\bf D46} (1992) 337.


\bibitem{nolr} 
A. Masiero, R. N. Mohapatra and R. Peccei, Nucl. Phys. {\bf B192} (1981) 66;\\
G. Branco and L. Lavoura, Phys. Lett.  {\bf B165} (1985) 327;\\
  J. Basecq, J. Liu, J. Milutinovic and L. Wolfenstein, Nucl. Phys. {\bf
B272} (1986) 145; \\
 J. F. Gunion, J. Grifols, A. Mendez, B. Kayser and F. Olness,
Phys. Rev. {\bf D40} (1989) 1546; \\
N. G. Deshpande, J. F. Gunion, B. Kayser and F. Olness,
Phys. Rev. {\bf D44} (1991) 837.

\bibitem{wu} L. J. Hall and S. Weinberg, \pr{D48} (1993) 79;\\
Y. L. Wu and L. Wolfenstein, \prl{73} (1994) 1762.

\bibitem{gabriela} G. Barenboim and J. Bernab\'eu, preprint  FTUV/96-9,
e-print hep-ph/9603379, to appear in Z. Phys. C.

\bibitem{bur} G. Buchalla, A. J. Buras and M. K. Harlander,
Nucl. Phys. {\bf B337},  (1990) 313; \\
G. Buchalla, A. J. Buras and M. K. Harlander,
Nucl. Phys. {\bf B355},  (1991) 305.


\bibitem{beall} G. Beall, M. Bander and A. Soni, Phys. Rev. Lett. {\bf 48}
(1982) 8484.


\bibitem{qmass} G. Rodrigo, preprint FTUV 95/30, e-print hep-ph/9507236
and references therein.

\bibitem{gut} K. S. Babu and R. N. Mohapatra, \prl{70} (1993) 2845; \\
L. Lavoura, \pr{D 48} (1993) 5440.

\end{thebibliography}
\end{document}